\documentclass[aps,prb,twocolumn,tighten,groupedaddress,amssymb,amsmath,showpacs]{revtex4-1}
\usepackage{dsfont}
\usepackage{bm}
\usepackage{bbm}
\usepackage[mathscr]{euscript}
\usepackage{amsmath}
\usepackage{amssymb}
\usepackage{amsthm}
\usepackage{amsfonts}
\usepackage{graphicx}
\usepackage{color}
\usepackage[colorlinks=true,urlcolor=blue,linkcolor=blue,citecolor=red]{hyperref}

\begin{document}


\title{Euler \& Lagrange versus Heisenberg \& Scr\"{o}dinger: Dynamical Pictures in Classical and Quantum Mechanics}


\author{M. Hossein \surname{Partovi}}
\email[Electronic address:\,\,]{hpartovi@csus.edu}
\affiliation{Department of Physics and Astronomy, California State
University, Sacramento, California 95819-6041}


\date{\today}

\begin{abstract}
Using quantum-classical analogies, we find that dynamical pictures of quantum mechanics have precise counterparts in classical mechanics.  In particular, the Eulerian and Lagrangian descriptions of fluid dynamics in classical mechanics are the analogs of the Schr\"{o}dinger and Heisenberg pictures in quantum mechanics, respectively.  Similarities between classical and quantum dynamical pictures are explored within the framework of the Koopman-von Neumann formalism.  These allow for a natural definition of various dynamical pictures in classical mechanics as well as the application of classical concepts to quantum dynamics.   As an illustration, we use the interaction picture to find the classical evolution of an ensemble of particles of equal initial momenta and arbitrary configuration density under the action of a constant force in one dimension.  As a second example, we discuss the extension of the ideas of sensitivity to initial conditions and chaos in classical mechanics to quantum mechanics.

\end{abstract}



\maketitle



\section{Introduction}
Similarities between quantum and classical mechanics are as striking as the profound differences that set them apart.  Thus the superposition principle of quantum theory which underlies the phenomenon of entanglement and much that is regarded as quantum strangeness is a drastic departure from classical concepts.  On the other hand, modern quantum mechanics, born as it was in the cradle of classical mechanics, bears unmistakable structural similarities to it.  This is most notable in the ground breaking contributions of Heisenberg and Dirac.\cite{Hei,Dir}  In particular, Dirac's discovery of the correspondence between the fundamental commutation relations of quantum dynamics and the Poisson bracket formulation of classical mechanics showed a remarkable resemblance in the canonical structures of the two dynamical regimes.   In the words of Dirac, ``Quantum mechanics was built up on a foundation of analogy with the Hamiltonian theory of classical mechanics."\cite{Dir}   Naturally, classical-quantum analogies have played an important role in broadening our understanding of the structure of quantum mechanics as well as providing invaluable insight in seeking new approaches.  An especially compelling instance of this is Feynman's discovery of the path integral formulation which was inspired by Dirac's remarks on the role of the Lagrangian in quantum mechanics.\cite{Fey}

This paper explores the analogies of classical and quantum mechanics, especially in connection with the structure of dynamical pictures in the two regimes.   Dynamical pictures are of course well known in quantum mechanics, the principal ones being the Heisenberg and  Schr\"{o}dinger pictures, with the Dirac picture an intermediate case between them.    That there exist analogs of quantum dynamical pictures in classical mechanics can be gleaned from the following observation.   All observable quantities in both dynamics may be formulated as average values of the quantities in question with respect to distribution functions characterising the state of the system, the density matrix
in quantum mechanics and the Liouville phase space density in classical mechanics.    The time evolution of the observable quantities, on the other hand, can originate entirely in the dynamical quantity (the Heisenberg picture), or entirely in the distribution function (the Scr\"{o}dinger picture), or in part from both (the Dirac picture).  Viewed in this manner, it is clear that entirely parallel considerations of dynamical pictures apply in classical dynamics, resulting in corresponding dynamical pictures.   Indeed we already have examples of the first two pictures in classical fluid mechanics, namely, the Lagrangian description as the counterpart of the Heisenberg picture and the Eulerian description the analog of the Scr\"{o}dinger picture, as we will be seen in detail below.   It will be recalled that the former references the initial conditions of all fluid particles and focuses on their individual paths during the flow, whereas the latter considers the velocity and acceleration of fluid particles as they pass through given points of space in the course of time.\cite{Lam}

In the following section we will set up the parallel formulations of classical and quantum dynamics discussed above, and derive the precise definition of dynamical pictures in classical dynamics.  In \S III we will use the interaction picture to treat the classical evolution of an ensemble of particles of equal initial momenta and arbitrary configuration density under the action of a constant force in one dimension.   In \S IV, we will discuss a precise formulation of sensitivity to initial conditions and the Lyapunov spectrum for quantum dynamics based on their counterparts in classical mechanics.  In \S V we will present a few concluding remarks.
\section{Parallel Formulations of Classical and Quantum Dynamics}
In comparing classical and quantum dynamics, it is convenient to use the Hilbert space formulation of classical mechanics provided by the Koopman-von Neumann formalism.\cite{KvN}    According to this description, the state of a dynamical system is represented by a wavefunction $\phi$ defined on a Hilbert space of functions of the canonical coordinates $(\mathbf{q}, \mathbf{p})$.   This wavefunction is postulated to obey the evolution equation
\begin{equation}
i\frac{\partial \phi(\mathbf{q}, \mathbf{p},t)}{\partial t}=\hat{L}\phi(\mathbf{q}, \mathbf{p},t),   \label{1}
\end{equation}
where  $\hat{L} =iL$, with $L$ is the standard Liouville operator
\begin{equation}
L={\sum}_{i} [\frac{\partial}{\partial {q}_{i}}H(\mathbf{q}, \mathbf{p})\frac{\partial}{\partial{p}_{i}}   - \frac{\partial}{\partial {p}_{i}}H(\mathbf{q}, \mathbf{p})\frac{\partial}{\partial {q}_{i}}].   \label{2}
\end{equation}
Here $H(\mathbf{q}, \mathbf{p})$ is the underlying classical Hamiltonian.

The classical phase space density associated with $\phi$ is then given by $\rho(\mathbf{q}, \mathbf{p},t)={\mid \phi \mid}^{2}$, which equals the probability density that the system has coordinates and momenta $(\mathbf{q}, \mathbf{p})$.    Using the postulated Eq.~(\ref{1}), we find that the resulting probability distribution obeys the standard Liouville equation:
\begin{equation}
\frac{\partial \rho(\mathbf{q}, \mathbf{p},t)}{\partial t}=[L,\rho(\mathbf{q}, \mathbf{p},t)].   \label{3}
\end{equation}

 The average value of any dynamical quantity $A(\mathbf{q}, \mathbf{p}) $ can now be expressed as
 \begin{equation}
<A>(t)=\int d\mathbf{q}d\mathbf{p}A(\mathbf{q}, \mathbf{p})\rho(\mathbf{q}, \mathbf{p},t) ,   \label{4}
\end{equation}
 or symbolically as
 \begin{equation}
<A>(t)=\mathrm{tr}[\hat{A}\hat{\rho}(t)].   \label{5}
\end{equation}

It is clear from Eqs.~(\ref{3}) and (\ref{4}) that the time dependence of $<A>(t)$ originates in $\hat{\rho}(t)$.   Thus the description given in Eqs.~(\ref{3}-\ref{5}) is in the Schr\"{o}dinger picture since the time dependence of average values originates in the quantity that represents the state of the system.

To construct the corresponding Heisenberg description, we first consider the particle trajectories $[\mathbf{Q}(t,{\mathbf{Q}}_{0},{\mathbf{P}}_{0}),\mathbf{P}(t,{\mathbf{Q}}_{0},{\mathbf{P}}_{0})]$ obeying Hamilton's equations of motion
\begin{equation}
i\frac{d (\mathbf{Q},\mathbf{P})}{d t}=-\hat{L}(\mathbf{Q},\mathbf{P}),   \label{6}
\end{equation}
and subject to the initial conditions $({\mathbf{Q}}_{0},{\mathbf{P}}_{0})$ at time $t=0$.    We can then relate the phase space density at time $t$ to its initial distribution using these trajectories:
\begin{equation}
\rho[\mathbf{Q}, \mathbf{P},t]= \rho[\mathbf{Q}_{0}(t, \mathbf{Q}, \mathbf{P}), \mathbf{P}_{0}(t, \mathbf{Q}, \mathbf{P}),0],   \label{7}
\end{equation}
where we have inverted the trajectory equations in writing the initial coordinates  $({\mathbf{Q}}_{0},{\mathbf{P}}_{0})$ in terms of their corresponding values at time $t$.    Equation (\ref{7}) is a consequence of Liouville's theorem expressing the constancy of the phase space density along a given trajectory.

At this point we can substitute the right-hand side of Eq.~(\ref{7}) in Eq.~(\ref{4}), change variables of integration to $({\mathbf{Q}}_{0},{\mathbf{P}}_{0})$ (remembering that the Jacobian equals unity due to Liouville's theorem), and arrive at
 \begin{align}
<A>(t)=&\int d{\mathbf{Q}}_{0}d{\mathbf{P}}_{0}A[\mathbf{Q}(t,{\mathbf{Q}}_{0},{\mathbf{P}}_{0}), \mathbf{P}(t,{\mathbf{Q}}_{0},{\mathbf{P}}_{0}),t]  \nonumber \\
& \times \rho[(\mathbf{Q}_{0}, \mathbf{P}_{0}) , 0].  \label{8}
\end{align}

Compared to Eq.~(\ref{4}), the representation in Eq.~(\ref{8}) has clearly shifted the origin of the time dependence of $<A>(t)$ to the dynamical variable while the phase space density is taken at the fixed initial time.   This is of course the Heisenberg picture of the time evolution of the system.   The time-dependent form of the dynamical variable $\hat{A}(t)$, on the other hand, is given by $A[\mathbf{Q}(t,{\mathbf{Q}}_{0},{\mathbf{P}}_{0}), \mathbf{P}(t,{\mathbf{Q}}_{0},{\mathbf{P}}_{0}),t]$, exactly as expected.

We note that Eq.~(\ref{8}) can be more formally obtained from Eq.~(\ref{5}) by observing that $\hat{\rho}(t)=\exp[-i\hat{L}t]\hat{\rho}\exp[i\hat{L}t]$, $ \hat{\rho}=\hat{\rho}(0)$:
\begin{align}
<A>(t)=&\mathrm{tr}[\hat{A}\hat{\rho}(t)]=\mathrm{tr}\{\hat{A}\exp[-i\hat{L}t]\hat{\rho}\exp[i\hat{L}t]\} \nonumber  \\
=&\mathrm{tr}\{\exp[i\hat{L}t]\hat{A}\exp[-i\hat{L}t]\hat{\rho}\}=\mathrm{tr}[\hat{A}(t)\hat{\rho}].   \label{9}
\end{align}
Note that, by applying Eq.~(\ref{6}), we can verify that the definition of $\hat{A}(t)$ resulting from this equation agrees with what we deduced above from Eq.~(\ref{8}).

It should now be clear that the above transformations are formally analogous to their quantum counterparts as the two regimes rest on similar canonical structures.

At this juncture we can consider fluid flow as an example and notice that Eq.~(\ref{8}) corresponds to the Lagrangian description whereby all dynamical quantities are averages over particle trajectories starting from an initial configuration.   Equation (\ref{4}), on the other hand, corresponds to the Eulerian description whereby the phase space density is considered a function of time, from which dynamical averages are obtained using time independent (equivalently, initial-time) expressions of those variables.

The foregoing discussion has established the two principal dynamical pictures in classical mechanics.   Evidently other equivalent pictures can similarly be defined.   We will consider Dirac's interaction picture in the following section.
\section{Interaction Picture in Classical Mechanics}
The interaction picture in classical mechanics is constructed in analogy with quantum mechanics.    Consider a splitting of $\hat{L}$ into two pieces,  $\hat{L}={\hat{L}}_{0}+{\hat{L}}_{I}$, corresponding to splitting the Hamiltonian into ``free'' and ``interaction'' parts, $\hat{H}={\hat{H}}_{0}+{\hat{H}}_{I}$.   Then, defining the interaction picture quantities as $\hat{\rho}_{I}(t)=\exp[i{\hat{L}}_{0}t]\hat{\rho}(t)\exp[-i{\hat{L}}_{0}t]$ and ${\hat{L}}_{I}(t)=\exp[i{\hat{L}}_{0}t]\hat{L}_{I}\exp[-i{\hat{L}}_{0}t]$, we find
\begin{equation}
i\frac{\partial {\hat{\rho}_{I}(t)}}{\partial t}={\hat{L}}_{I}(t){\hat{\rho}}_{I}(t),   \label{10}
\end{equation}
while the expectation value of $A$ is given by
\begin{equation}
<A>(t)=\mathrm{tr}[{\hat{A}}_{I}{\hat{\rho}}_{I}(t)],   \label{11}
\end{equation}
where $\hat{A}_{I}(t)=\exp[i{\hat{L}}_{I}(t)]\hat{A}\exp[-i{\hat{L}}_{I}(t)]$.   As expected, these results are analogous to their quantum counterparts.

To complete the analogy, we can rewrite Eq.~(\ref{10}) by defining an evolution operator $\hat{U}(t)$ such that $ {\hat{\rho}_{I}(t)}=\hat{U}(t){\hat{\rho}_{I}(0)}$.   Then
\begin{equation}
i\frac{\partial {\hat{U}(t)}}{\partial t}={\hat{L}}_{I}(t){\hat{U}}(t),   \label{11.1}
\end{equation}
subject to $\hat{U}(0)=1$.    In strict analogy with quantum mechanics, we can construct a perturbative solution to Eq.~(\ref{11.1}) using Dyson's formula:\cite{Dys}
\begin{equation}
\hat{U}(t)={T}\exp [-i\int_0^t d\tau {\hat{L}}_{I}(\tau)],   \label{11.2}
\end{equation}
where ${T}$ is the time ordering operator.

To see an illustration of the foregoing formulation, consider the case of motion in one dimension with $H(q,p)={{p}^2}/{2m}+V(q)$.    Then ${\hat{L}}_{0}=-i(p/m)\partial /\partial q$ and ${\hat{L}}_{I}=i{V}^{'}(q)\partial /\partial p$.    We also have ${\hat{L}}_{I}(t)=\exp[i{\hat{L}}_{0}t][i{V}^{'}(q)\partial /\partial p]\exp[-i{\hat{L}}_{0}t]$, which leads to\cite{dis}
\begin{equation}
{\hat{L}}_{I}(t)=i{V}^{'}(q+pt/m)[\partial/\partial p -(t/m)\partial/\partial q].   \label{12}
\end{equation}
As a simple application, let us consider an initial distribution of particles of density $f(q)$, all moving with momentum ${p}_{0}$, and subject to a constant force $F=-V'(q)$ starting at $t=0$.     Then the initial density is given by $f(q)\delta(p-{p}_{0})$, so that $\hat{\rho}_{I}(t)=\exp[i\hat{{L}_{0}}t]\exp[-i\hat{{L}}t]f(q)\delta(p-{p}_{0})$.

Next, we use the Baker-Hausdorff identity $\exp(A+B)=\exp(A)\exp(B)\exp(-[A,B]/2)$, where $\exp([A,B])$ commutes wit both $A$ and $B$, to find that
\begin{align}
\exp[i{\hat{L}}_{0}t] \exp[-i\hat{L}t]&=\exp(-Ft\partial/\partial p) \nonumber \\
&\times\exp[(F{t}^2/2m)\partial / \partial q],  \label{13}
\end{align}

which we apply to the evaluation of ${{\rho}}_{I}(q,p,t)$:
\begin{align}
&{\rho}_{I}(q,p,t)=\exp(-Ft\partial / \partial p)\exp[(F{t}^{2}/2m)\partial/\partial q]f(q)  \nonumber  \\
&\times \delta(p-{p}_{0})=f(q+\frac{F{t}^{2}}{2m})\delta[p-({p}_{0}+Ft)].   \label{14}
\end{align}
Finally, we can find ${\rho}(q,p,t)$ from Eq.~(\ref{14}) using $\hat{\rho}(t)=\exp(-{\hat{L}}_{0}t)\hat{\rho}_{I}(t)$:
\begin{equation}
{\rho}(q,p,t)=f(q+\frac{pt}{m}+\frac{F{t}^{2}}{2m}) \delta[p-({p}_{0}+Ft)],  \label{15}
\end{equation}
which corresponds to the phase space density displaced from position $q$ to $q+{pt}/{m}+{F{t}^{2}}/{2m}$ with momentum ${p}_{0}+Ft$, exactly as expected under the action of a constant force.

The above calculations serve to illustrate the formal structure of the interaction picture in classical mechanics.   As an application of the classical-quantum dynamical picture analogy that shows its utility in dealing with difficult questions, we will discuss the precise meaning of sensitivity to initial conditions and chaos in quantum dynamics in the following section.
\section{Sensitivity to Initial Conditions in Quantum Dynamics}
Classical chaos is a long term dynamical instability which causes initially close trajectories to diverge exponentially and eventually lose memory of their initial conditions.\cite{EcR}   It is ubiquitous among nonlinear systems of more than one degree of freedom.   Chaotic behavior is measured in terms of the rate of trajectory divergence mentioned above, and when present is characterized by an effectively random behavior for sufficiently long times.   Because the quantitative measure of chaos is based on classical trajectories, it cannot be directly applied to quantum dynamics.
As will be seen below, it turns out that the above-mentioned instability is absent in quantum dynamics.\cite{Abs}   On the other hand, there are a number of characteristic behaviors exhibited by quantum systems whose classical versions are chaotic.   These are a subject of active study often referred to as ``quantum chaos.''\cite{Gut,CaS}

The aim of this section is to demonstrate that the correspondence detailed in the previous sections allows a precise measure of chaos in quantum dynamics by analogy to classical dynamic.

\subsection{Chaos in Classical Dynamics}
We start by outlining the quantitative definition of chaos in classical mechanics.\cite{EcR,Red,Can}    Consider a classical system of $N$ degrees of freedom described by $2N$ canonical variables $\{
{q}_{i},{p}_{i} \}$, $i=1, \ldots,N$ and Hamiltonian $H({\bf q},{\bf p})$.   The trajectory divergence behavior of a system is best measured by following the evolution of an infinitesimal $2N$-dimensional sphere in phase space initially centered at point $({\mathbf{q}}_{0},{\mathbf{p}}_{0})$.   Such a sphere will be distorted in the course of time, stretching in some directions and contracting in others, while maintaining a fixed $2N$-dimensional.   The details of these deformations are fully conveyed by the Jacobian of the flow, namely the $2N \times 2N$, nonsingular block matrix
\begin{equation}
{\cal T}(t){=} \left( \begin{array}{cc}
 {\partial \mathbf{q(t,{\mathbf{q}}_{0},{\mathbf{p}}_{0})}}/{\partial {\mathbf{q}}_{0}} &{\partial \mathbf{q(t,{\mathbf{q}}_{0},{\mathbf{p}}_{0})}}/{\partial {\mathbf{p}}_{0}}    \\
{\partial \mathbf{p(t,{\mathbf{q}}_{0},{\mathbf{p}}_{0})}}/{\partial {\mathbf{q}}_{0}} &{\partial \mathbf{p(t,{\mathbf{q}}_{0},{\mathbf{p}}_{0})}}/{\partial {\mathbf{p}}_{0}}    \end{array} \right),  \label{16}
\end{equation}
where $[{\mathbf{q(t,{\mathbf{q}}_{0},{\mathbf{p}}_{0})}},{\mathbf{q(t,{\mathbf{q}}_{0},{\mathbf{p}}_{0})}}] $ is the phase-space trajectory that starts from $({\mathbf{q}}_{0},{\mathbf{p}}_{0})$.  We shall refer to this Jacobian as the {\it sensitivity} matrix.   It has a unit determinant by Liouville's theorem.

A long-time exponential growth in any of the singular values of this sensitivity matrix signals chaotic behavior.   Thus we consider the Hermitian matrix $\ln({\mathcal{T}}^{\dag}{\mathcal{T}}/2t)$, whose eigenvalues as $t\rightarrow \infty$ constitute the Lyapunov spectrum, or indices, for the system; the dagger here represents the transpose of a matrix.   These indices represent exponential growth rates in certain phase-space directions, and they occur in pairs of equal magnitude and opposite signs.  For a non-chaotic system, all indices vanish, while for a chaotic system, a number of indices ($\leq N-1$) can be positive.   The sum of such positive indices, $h$, is known as the metric or KS entropy, after Kolmogorov and Sinai,\cite{Bil} and is an important information theoretical invariant quantity associated with a dynamical system.  Thus a system is chaotic if, and only if, $h >0$.   As mentioned above, the long-term behavior of a chaotic system is effectively random even though the system is governed by fully deterministic equations of motion.   How such an outcome is possible is best understood in terms of the concept of \textit{algorithmic information content}, which we discuss next.

In order to characterize the information content of a system, Solomonoff, Kolmogorov, and Chaitin, independently and nearly simultaneously, introduced the idea of algorithmic information content, defined as the length of the most economical description of the system.\cite{SKC}  This seemingly ambiguous characterization can be made mathematically precise, and as such provides a rigorous definition of a random sequence as one which does not allow a significantly shorter description.   Equivalently, a random sequence is one that cannot be algorithmically compressed.   As an example, we note that the seemingly random digits in the decimal (or binary) representation of the number $\pi$ are very highly compressible and far from random since this irrational number (hence also its various representations) can be defined by a short sentence.   In the case of an infinite sequence, its \textit{algorithmic complexity} $k$ is defined to be the ratio of the shortest description to length of its subsequences in the limit of longer and longer subsequences.   If $k>0$, the sequence is said to be complex and is considered random.   In particular, such a sequence cannot be defined by a finite one.

The relevance of the above ideas arises from a profound theorem that shows that the output of a dynamical system is complex if, and only if, its metric entropy $h$ is positive, in which case its algorithmic complexity and metric entropy are equal, i.e., $h=k$.\cite{ALY}   In other words, the long-time output of a chaotic system (with $k>0$) is indistinguishable from random.   In practical terms, this outcome can be understood by considering the fact that any initial conditions can only be specified with finite accuracy, say to 100 decimal places, which will progressively be eroded in the course of time (due to the sensitivity condition) and totally lost in finite time (due to the exponential character of the sensitivity to initial conditions), resulting in output which is unrelated to the initial conditions.

It is worth emphasizing that the foregoing characterization of a chaotic system is tied to its long-time behavior and is best understood in information theoretical terms.    Note also that the definition of chaos in terms of the complexity of its dynamical output is general and directly applicable to quantum systems.

\subsection{Sensitivity Operator in Quantum Mechanics}
Using the quantum-classical analogies described above, we will next develop a quantum counterpart of the sensitivity matrix for quantum dynamics.    First we note that the definition of the latter in Eq.~(\ref{16}) is in the Heisenberg representation of classical mechanics inasmuch as the time dependence is carried by the dynamical variable.    Thus we construct the analogous object in quantum mechanics by applying Dirac's correspondence rules to Eq.~(\ref{16}):
\begin{equation}
\hat{{\cal T}}(t){=}\frac{-i}{\hbar} \left( \begin{array}{cc}
[\hat{\mathbf{q}}(t),{\hat{\mathbf{p}}}_{0}]&-[\hat{\mathbf{q}}(t),{\hat{\mathbf{q}}}_{0}]    \\
{[\hat{\mathbf{p}}(t),{\hat{\mathbf{p}}}_{0}]}&-[\hat{\mathbf{p}}(t),{\hat{\mathbf{q}}}_{0}]    \end{array} \right).  \label{17}
\end{equation}
We shall refer to this matrix of operators as the \textit{sensitivity operator}.    The elements of this object constitute a set of $4{N}^{2}$ Hermitian operators, so that its expectation value for a quantum state $\hat{\rho}$, i.e., $\textrm{tr}[\hat{\cal T}(t) \hat{\rho}]$, results in a real sensitivity matrix ${\cal T}(t)$ which is the counterpart of the classical object.   The Lyapunov indices can then be calculated as in the classical case.  In particular, any exponential growth in time occurring in any of the components of this matrix signals chaotic behavior.   As seen in the following, such behavior does not occur in quantum mechanics.

Consider a typical element of $\hat{\cal T}(t)$, say $\frac{-i}{\hbar}[{\hat{q}}_{i}(t),{\hat{p}}_{j}(0)]$, and a quantum state $\hat{\rho}$.   Let ${\bar{q}}_{i}(t)=\textrm{tr}[{\hat{q}}_{i}(t) ]\hat{\rho}]$ and  ${\bar{p}}_{j}(0)=\textrm{tr}[{\hat{p}}_{j}(0) \hat{\rho}]$ be the corresponding mean values of the canonical coordinates.   We also recall the (generalized) Heisenberg inequality ${|\textrm{tr}[\hat{A},\hat{B}]\hat{\rho}|}^{2}\leq 4 \textrm{tr}[{\hat{A}}^{2}\hat{\rho}]\textrm{tr}[{\hat{B}}^{2}\hat{\rho}]$, which we apply with $\hat{A}={\hat{q}}_{i}(t)-{\bar{q}}_{i}(t)$ and $\hat{B}={\hat{p}}_{j}(0)-{\bar{p}}_{j}(0)$.    The result is
\begin{equation}
|{\hat{\cal T}}_{ij}(t)|=|\textrm{tr}[{\hat{q}}_{i}(t)-{\bar{q}}_{i}(t),{\hat{p}}_{j}(0)-{\bar{p}}_{j}(0)]\hat{\rho}| \leq \frac{2}{\hbar}{\delta {\hat{q}}_{i}(t)} {\delta {\hat{p}}_{j}(0)}, \label{18}
\end{equation}
where ${\delta {\hat{q}}_{i}(t)}$ stands for the square root of the variance of the observable $\hat{q}_{i}(t)$.

Inequality (\ref{18}) establishes the lack of unbounded growth, exponential or otherwise, in the sensitivity operator for bounded quantum states.   It is worth pointing out here that sensitivity to initial conditions and chaos are concepts that are appropriate to bounded systems.    Since the right-hand side of (\ref{18}) is related to variances of bounded canonical coordinates, we conclude that the sensitivity operator in quantum mechanics is bounded and cannot exhibit sensitivity to initial conditions.

While there are other indications such as the discrete nature of the spectra of bounded quantum systems suggesting the same conclusion as above, we note the great generality of the foregoing reasoning and its close connection with the original definition of chaos in classical mechanics.
\section{Concluding Remarks}
Dynamical picture similarities are part of the landscape of quantum-classical analogies and play an important practical and conceptual role in extending our understanding of quantum mechanics.   These analogies notwithstanding, there are important structural differences between quantum and classical dynamics.   One such difference is evident from our discussion of sensitivity to initial conditions.  Considering the quantum-classical transition, how do we reconcile the lack of such sensitivity for a quantum system, where $\hbar \neq 0$, with its emergence as $\hbar \rightarrow 0$ leads to its chaotic classical counterpart?    The answer lies in the fact that chaos and sensitivity to initial conditions are long-time properties, and require the $t \rightarrow \infty$ limit which may not commute with the $\hbar \rightarrow 0$ limit.   Indeed the so-called quantum kicked rotor,\cite{CaS} whose classical version exhibits chaos, does appear to be sensitive to initial conditions for a finite period of time.   Its long-time behavior, however, is perfectly regular.  This example underscores the key role of the $t \rightarrow \infty$ limit in characterizing chaotic behavior.  More generally, it serves as a warning that, despite strong similarities between quantum and classical dynamics, the transition across their boundary is anything but smooth and must be handled carefully.

{}


\begin{thebibliography}{}
\bibitem{dis} Here and in the following we use the basic property of a displacement operator according to which $\exp(a\partial/\partial x)f(x)=f(x+a)$, where $a$ is independent of $x$.
\end{thebibliography}
\end{document}